\documentstyle[12pt,epsf]{article}
\def\be{\begin{equation}}
\def\ee{\end{equation}}
\def\bea{\begin{eqnarray}}
\def\eea{\end{eqnarray}}
\def\ba*{\begin{eqnarray*}}
\def\ea*{\end{eqnarray*}}
\begin{document}

\newcommand{\sheptitle}
{Birth of the Universe from the Multiverse }
\newcommand{\shepauthor}
{ Laura Mersini-Houghton}
\newcommand{\shepaddress}
{ Department of Physics and Astronomy, UNC-Chapel Hill, NC, 27599-3255, USA}

\begin{titlepage}
\begin{flushright}
%hep-ph/yymmnnn\\
%SNS-PH/02-04\\
\today
\end{flushright}
\vspace{.1in}
\begin{center}
{\large{\bf \sheptitle}}
\bigskip \medskip \\ \shepauthor \\ \mbox{} \\ {\it \shepaddress} \\
\vspace{.5in}
%{\bf Abstract}

\bigskip \end{center} \setcounter{page}{0}

\newcommand{\shepabstract}
\shepabstract

%\baselineskip 24pt

%\maketitle

%%%%%%%%

{It is fair to say that the deepest mystery in our understanding of nature is the birth of our universe. Much of the dilemma over the last decades comes from the extraordinarily small probability that the universe started with the high energy Big Bang as compared to the chance of nucleating any other event. How can Big Bang cosmology be $10^{10^{123}}$ times less likely than nucleating the present cold universe, while accumulating such exquisite agreement with astrophysical data? Why don't we see the other nucleations that, if left to chance, seem to overwhelmingly outnumber us? Here I discuss the point of view that the selection of the initial conditions can be meaningfully addressed only within the framework of the multiverse and that the reason why Big Bang inflation was preferred over other events lies in the quantum dynamics of the landscape of the initial patches. The out-of-equilibrium dynamics selected the 'survivor' universes be born at high energies and the 'terminal' universes at low energies. I briefly review the testable predictions of this theory, in particular the giant void observed in 2007. The second part focuses on the extended framework, in particular a set of postulates needed for defining the multiverse.}
%\begin{abstract}
%\end{abstract}
%{\Large \bf Equation of state }

\begin{flushleft}
\hspace*{0.9cm} \begin{tabular}{l} \\ \hline {\small Email:
mersini@physics.unc.edu} \\

\end{tabular}
\end{flushleft}

\end{titlepage}
%%%%%%%%%%%%%%%%%%%%%%%%%%%%%%%%%%%%%%%%%%%%%%%%%%%%%%%%%%%%%%%%%%%%%%%%%%%%%%%%%%%%%

\section{The Puzzle of the Initial Conditions}

At one time or another every human has looked up at the skies and wondered how did it all start. Modern cosmology provided the answer: our universe was born with Big Bang inflation. While the exact energy scale $\Lambda$ at which this event ocurred is not known, it is widely believed that the universe was at extremely high energies at the instance of the Big Bang, possibly at Grand Unified Scale, $GUT$ ($10^{16}$ GeV), or higher. Then as inflation propelled the universe into an accelerated expansion, the universe grew big and cooled down. Primordial fluctuations of this inflaton field then seeded the observed structure of the universe. This is the beauty of the inflation paradigm, it provides such a simple picture for such a complex and intricate system. The exquisite agreement of the inflationary scenario with the vast amount of astrophysical data accumulated over the last few decades makes this paradigm even more impressive. But what is really amazing is how unnatural inflation is when considering that the probability for the high energy Big Bang inflation to occur, is extraordinarily small by standard estimates. And here lies the puzzle: what selected such extremely unlikely initial conditions for the birth of our universe?

The probability $P$ depends exponentially on the entropy $S$ of the initial patch at Big Bang, $P = e^S$, (where the entropy is a quantity that measures the amount of disorder or information contained in the universe). But since for systems in {\it thermal equilibrum}, $S$ is inversely proportional to the energy scale $\Lambda$, then the probability is exponentially suppressed with the energy at the time of the Big Bang, $P = e^{1/\Lambda}$. Thus the higher the energy scale at which Big Bang switches on, the less likely it is for the universe to nucleate from such an event. In fact, if our universe started at $GUT$ scale, Roger Penrose estimated that such an event has only one chance in $10^{10^{123}}$ possibilities. At first, this estimate seems to imply that the birth of the universe was a very special event indeed. It also implies that since other events are overwhelmingly more likely to nucleate then they should be present around us today in abundance and worse, these events should outnumber us. This latter paradox is known as the Boltzmann Brains (BB's) paradox. A third problem stemming from the puzzle of the selection of the initial conditions is the problem of the thermodynamic arrow of time: we observe that time flows in a definite orientation from past to future, an orientation which, according to the second law of thermodynamics (an $empirical$ law by itself) is determined by the direction of the entropy increase. The trouble here is that the universe had to start from a low entropy state in order to have the observed arrow of time. But the price to pay is that a low entropy initial state suggests an exponentially small chance for that event to have occurred. We have not seen any BB's around, we observe that time flows from past to future. But we also know that the theory of Big Bang inflation, known as {\it standard cosmology}, has passed all observational tests. Yet it appears so unnatural and unlikely when faced with the question: why did we start with such an extremely ordered initial condition and, what selected this special state?

The deep mystery of the initial conditions and the paradoxes that standard estimates produce, are probably indicating that the problem lies with our understanding and in the theoretical framework we are using, rather than in flaws of inflation or anthropic designs.
As is often the case in physics, progress towards probing one mystery of nature results in trading-off the old set of questions for new ones that are usually deeper and more complex. Inflation addressed our questions about the observed flatness, homogeneity and structure in the universe. But it produced a new and more severe mystery, that of the selection of its initial state,which for high energy inflation seems to be the most unlikely state the universe could pick, one that is far less likely than starting with a big cold universe with all the internal structure of our current universe.

Understanding the birth of our universe is the most important mystery in our understanding of nature. I have been advocating that the only way to meaningfully address this puzzle is by extending our physical theories within a multiverse framework; by dropping all the unwarranted assumptions of thermal equilibrum and ergodicity and, by investigating the quantum dynamics of all the degrees of freedom in this ensemble of initial patches, especially the dynamics of gravitational degrees of freedom. As I describe below, by doing so yields a much more reasonable answer for the chance of starting the universe with high energy inflation. An exciting prospect described further below, is that this approach makes predictions that can be tested by our present and near future data, thereby giving this theory the potential to be falsified.

\section{Why did the  Universe Start in such an Extremely Ordered State?}

We talk of entropy as a measure of disorder. Thus the lower the entropy the more ordered the system. The usual example given is that a broken glass is in a state of higher disorder and higher entropy as compared to a glass that remains intact. Disorder is a fuzzy concept to quantify and an intuitive grasp of entropy is not sufficient. Yet, understanding the selection of the initial conditions for our universe is equivalent to a deeper understanding of the entropy of the universe, both quantitatively and qualitatively. So, we need to dissect this mystery into more transparent ingredients. To do so, it is helpful to look at the concept of entropy in a different way, in terms of information contained in this disorder which counts the number of states available to the disordered system. Information in a system is carried in correlations among its degrees of freedom. Entropy measures the amount of information, (equivalent to disorder), contained in a system. Correlations can be calculated mathematically in a precise manner, and yet we are on shaky grounds when defining the entropy in our universe and subsequently the probability for such a universe to exist. So, although we know how to estimate correlations, the mystery of the entropy and initial conditions of the universe boil down to: which correlations are we to include in the entropy of the universe?

As described in the previous section the general estimate of the entropy of the universe was based on the 'black box' model. This means we make a major assumption that degrees of freedom in the universe are in perfect thermal equilibrum, and that the counting of entropy internally reflects our ignorance of the information outside which is cut off by the walls of the box,and which, in this case the horizon of the universe. In the 'black box' approach the correlations among the internal and external degrees of freedom are assumed to be zero.  

In 2005 I took the point of view that, in order to get a sensible answer, we need to incorporate the dynamics of gravitational degrees of freedom and study the problem of the entropy of the universe as a {\it non-equlibrum} system within the framework of the extended physics of the multiverse, \cite{land1,land2,laurareview, laurarich1, laurarich2}.

There are good reasons to take this approach, which I'll briefly account for here:

{\bf i.}   Our previous assumptions in studying the initial conditions for equilibrum and ergodicity, can not be justified. Based on these assumptions we are led to the expression $P=e^S$ followed by a series of paradoxes when trying to explain the birth of our universe. Therefore it is best we do not make these assumptions without checking whether they hold before using them. As was shown in \cite{laurarich1,laurarich2} the very early universe is far out of equlibrum and the phase space is not ergodic due to being cleaned out of the low energy 'terminal' universes.

{\bf ii.}  Gravitational degrees of freedom play an important role in the overall dynamics, especially in the very early universe when this interaction is non-negligible. To illustrate this point let us think of the gravitational dynamics of the vacuum energy $\Lambda$ that gives rise to Big Bang inflation. As is well known the tendency of any system is to try and reach equilibrum by maximizing entropy. But as matter degrees of freedom tend to equlibrum by trying to pull the initial patch to a ('black hole') crunch, the gravitational degrees of freedom contained in $\Lambda$ tend to equlibrum by trying to expand that initial patch to infinity. Since any realistic system of cosmology will contain both matter and gravitational degrees of freedom, then their opposing tendencies in this 'tug-of-war' on the initial patch guarantee two things: first, that this system is far from being in equlibrum; and second, that their dynamics provides a (super-)selection rule for the initial conditions of the universe by leading to 'survivor' and 'terminal' universes. If $\Lambda$ 'wins' over matter the initial patch grows giving rise to a {\it 'survivor' universe}, which is a physically relevant entity. But if the pull of matter is stronger than the effect of $\Lambda$ on the expansion of the universe, then the initial patch can not grow and contracts, resulting in a {\it 'terminal' universe}. Thus the selection of universes that start with high a energy scale for $\Lambda$ is naturally driven by the quantum dynamics of gravity. This ingredient, the quantum dynamics of gravitational degrees of freedom, was not included in the previous attempts at addressing the selection of the initial conditions. As is often the case in science, what we missed out is highly significant.

{\bf iii.} Another important aspect in this radically new way of thinking is the need to extend our physical theories within a multiverse framework. The dominant expectation so far for the ,(yet unknown), theory of quantum gravity (QG) has been the {\it 'reductionist'} hope that relies on QG producing one unique solution that resembles the general features of our universe. We have failed so far on this view. Au contraire, as described in more detail in Sec. $4-6$, the success of three major theories in physics has taught us otherwise.That the three different and important theories, quantum mechanics, string theory and inflation, all predict the existence of the multiverse is, I believe, hardly coincidental. How else can we ask the question: why did we start with these initial conditions, without implying, as compared to what other choices? If there is only one sample available to choose from then there is no point asking: why did we end up with this sample. We are guaranteed in that case that we can not end up with anything else but this sample. This 'reductionist' scenario does not allow any selection questions since it has already assigned the probability weights to be $1$ for the sample and $0$ for any other option. A question about selection thus makes sense only when we have a pool of 'universe samples' to choose from, (a.k.a a multiverse). So far I have made no distinction in relation to the question: is the multiverse a mathematical or a physical entity. Issues related to the structure and the embedding spacetime(s) of the multiverse are discussed in Sec.$4-6$. But the important argument I would like to present here is that the existence of the multiverse must be expected from the underlying fundamental theory or else we can not meaningfully address the mystery of the initial conditions. In this {\it 'meritocratic'} view, the extension of our current cosmology within a multiverse framework becomes an extension of the Copernican principle to nature. In Sec.$4-6$, in order to discuss the ontology of the multiverse I propose to apply two principles:\\
- the principle of 'No Perpetual Motion' as a criterion for the parameter of time; and,\\
- the principle of 'Domains Correlations' as a criterion for determining the background spacetime in which the multiverse is embedded.

\subsection{Selection of the Initial Conditions for our Universe on the Landscape}

Having now laid out the main ingredients, we are ready to apply this dynamic approach to the selection of the initial conditions for our universe within the multiverse framework.

 The landscape of string theory \cite{landscape,kklt,KKLMMT,bousso,douglas,ashok} will be our working model for providing a physically motivated example of the multiverse space, for the following reasons: every landscape vacua hosts a potential birth place for a universe. Therefore the collection of all string vacua solutions that comprise the landscape, is an ensemble of possible universes \cite{laurarich1,laurarich2,land1,land2,avatars1,avatars2}. We can think of the landscape of string theory as the energy profile $V(\phi)$ of some string moduli field $\phi$, which plays the role of a collective coordinate. The internal degrees of freedom in each vacua site, are centered around the mean vacua field $\phi_j$ with vacuum energy $\Lambda_j = V(\phi_j)$, and can be thought as perturbations around the mean value of the field $\phi$. In an important work, Michael Douglas and Frederik Denef \cite{douglas} studied the structure and distribution of vacua in the landscape. I will use their model for the landscape structure in what follows below, including the distribution of the internal degrees of freedom in each vacua, (see \cite{laurarich1,laurarich2} for the technical details). The distribution they found for the vacua energies on the landscape is very similar to randomly disordered condensed matter systems and quantum dots, as discussed in \cite{laurarich1}, known as the universality class of CI-type \cite{altland}. In the case of the landscape the number of vacua $'j'$ is expected to be in the order of $10^{500}$.

 I will treat the problem quantum mechanically and make the assumption that quantum mechanics is a fundamental theory of nature, that is, that this theory is valid at all energy scales. Quantum mechanics can be embedded into the landscape multiverse by allowing the wavefunction of the universe \cite{hh, DeWitt, Vilenkin} to propagate on the landscape of string theory, which was proposed in \cite{land1,land2}. The solutions to the 'Schrodinger'-like equation, which in quantum cosmology is known as the Wheeler DeWitt (WdW) equation \cite{DeWitt, Vilenkin}, can be calculated by analogy with the above condensed matter systems, using random matrix theory. The important caveat worth emphasizing is the subtlety related to the time parameter: the intrinsic time in the wavefunction should be interpreted as fundamental, like it is in quantum mechanics, rather than born out of WdW solutions. 

The WdW equation is defined on a two dimensional ($2-dim$ hereon), minisuperspace parametrized by two variables, $(a,\phi)$: the scale factor $a(t)$ of universes with spatially flat $3-dim$ geometries, given by the metric: $ds^2 = dt^2 - a^2 dx^2$, where $x$ stands for the $3-dim$ space variable and $t$ for time; and, by the moduli field $\phi$ of the landscape, with energy $V(\phi)$, on which the wavefunction of the universe propagates. Then the WdW equation states that the hamiltonian operator ${\cal H}$ on the minisuperspace, which characterizes the total energy of this 2-variable system, when acting on the wavefunctional of the universe $\Psi[a,\phi]$, must give zero energies. (For this reason the WdW equation is often referred to as the 'constraint equation'). The WdW equation is:

\begin{eqnarray}
& &{\hat {\cal H}}\Psi(a,\phi) = 0 ~{\rm with} \nonumber \\
& &\hat{{\cal H}}=\frac{1}{2e^{3\alpha}}\left[\frac{4\pi}{3M_p^2}
\frac{\partial^2}{\partial\alpha^2}-
\frac{\partial^2}{\partial\phi^2}+e^{6\alpha}V(\phi)\right].
\label{2}
\end{eqnarray}
For mathematical simplicity, here the scale factor $a$ has been written as $a=e^{\alpha}$ and the energy around the vacua sites approximated by the expression $V(\alpha, \phi) = e^{6\alpha}m^2 \phi^2 + e^{4\alpha}\kappa, \kappa =0,-1$ for flat or closed universes, and $M_p$ is the reduced Planck mass, (which is related to the usual Planck mass $m_p$ by absorbing a factor of $8\pi$, $8\pi M_p^{2} = m_p^{2}$, and is often used in equations for the simplicity of short-hand notation). The ``mass''-like parameter, $m^2$, corresponds to the curvature of the vacua, $ m^2 \simeq d^2 V / d \phi^2 $, vacua with energy $\Lambda \simeq m^2 \phi^2 / 2 $. The hat symbol denotes the quantum mechanical operator. From Eqn.1 it can be seen that indeed the hamiltonian ${\cal H}$ contains the kinetic and potential energies for both variables, the scale factor $a$ corresponding to the $3-$ geometries and the moduli $\phi$ corresponding to the landscape vacua energies. 

The probability distribution for the wavefunction of the universe solutions is found by the quantum mechanical expression of the norm squared, $P=|\Psi[a,\phi]|^2$. Calculated in \cite{land1,land2} this expression shows that the most probable universe is found to sit at vacua with the minimum nonnegative, i.e. zero, energies. Clearly this is not correct and we know that our universe started at high energies. The reason for this incorrect result lies with the limitations of the WdW equation which is not sufficient for treating the quantum dynamics of the system \cite{laurarich1,laurarich2}. We need to incorporate in our investigation the issue of how these 'universe' solutions decohere (decouple) from each other at later times when they undergo a quantum to classical transitions. 

Decoherence is induced by the 'environment' or 'bath' during the process of measuring the 'system' \cite{qcreview}, thereby triggering the quantum to classical transition for the $(3+1)-dim$ worlds and the decay of entanglement among the different worlds \cite{zeh}. We thus need to identify 'the bath' and 'the system' for our problem. Decoherence \cite{qcreview,zeh,zurek} in this problem is taken into account by including the effect of the 'bath' consisting of superhorizon sized wavelength massive fluctuations, into the 'system', which is comprised of degrees of freedom inside the horizon of the universe.

The wavepacket with the $(3+1)-dim$ geometry sitting in one of the landscape vacua is 'the system', ( which in real spacetime eventually becomes our universe). But the 'measurer', the environment, needs to be chosen carefully such that, when it interacts with the system, it does not interfere with the outcome of the observable that it is measuring. A candidate for the 'bath', are  the degrees of freedom of the very long massive perturbations, with superhorizon size wavelengths $\lambda_n$. These perturbation modes are coupled gravitationally to the 'system' which consists of all degrees of freedom inside the horizon $H$. Since the strength of coupling goes as $G_{N}/{\lambda}$, with $G_N$ being the Newton's constant, then for really large wavelengths $\lambda_n \ge l_H$, where $l_H$ is the horizon size, the strength of coupling of these environment modes to the system is very small, guaranteeing a very weak coupling of these modes with the 'system' as they 'watch' the system. Superhorizon sized perturbations are thus our chosen 'environment' in which the 'system' is immersed. But there is an infinite number of these fluctuations $f_ n$ in the 'bath'. The 'bath' consists of all fluctuations with $\lambda_n$ in the range from $l_H$ to infinity, ( where $n$ is the label that counts the modes), and energies: ${\cal H}_n = - \frac{\partial^2}{\partial f_n^{2}} + {\cal U}_n f_n^{2} $, calculated in \cite{gibbonshawking}, where the first term represents the kinetic energy and the second term ${\cal U}_n f_n^{2}$ represents the potential energy of these $f_n$ modes with $U_n$ similar to a mass term \cite{kiefer}. The total contribution of the 'bath' given by the sum over all its modes $\Sigma_n {\cal H}_n$, backreacts on the geometry of the 'system' by perturbing the variables $a,\phi$ in the WdW equation. Let's now add the sum $\Sigma {\cal H}_n$ of the total contribution of these perturbation modes with $\lambda_n \ge l_H$ into the WdW equation, \cite{laurarich1,laurarich2}. This new term added to the WdW hamiltonian ${\cal H}_0$, represents the backreaction of the 'bath' onto the 'system ' \cite{qcreview,zeh}. With the inclusion of the backreaction term, the WdW equation becomes a 'Master' equation \cite{laurarich1,avatars1,avatars2}

\begin{equation}
\hat{\cal H}_0 \Psi[a,\phi]  = -\Sigma \hat{\cal H}_n \Psi[a,\phi],
\end{equation}

The Master Equation is now not constrained to zero energies due to the new backreaction term ${\cal H}_n$. The interesting part is that during inflation, the backreaction term is negative, playing the role of a negative mass-squared term in the Master Equation \cite{kiefer}. The significance of a negative mass is that it creates gravitational instabilities in the solutions. Thus the backreaction of superhorizon sized fluctuations creates a gravitational instability for the wavefunction of the universe by counterbalancing the effect of the vacuum energy $\Lambda \simeq m^2 \phi^2$ in driving the expansion of the initial $(3+1)-dim$ world. When does this instability occur? If the $\Lambda$ term is larger than $\Sigma {\cal H}_n$ then the overall mass given by the sum of the two terms is positive and the instability does not occur. That means that solutions for the wavefunction of the universe exist and a universe is born for each vacua on the landscape that has a vacuum energy $\Lambda$ larger then the backreaction term. On the other hand, all the vacua with energies $\Lambda \le \Sigma {\cal H}_n$ host 3-geometries/worlds that can not survive and can not expand due to the gravitational instabilites induced on these initial patches from the strong backreaction term. In this case, mathematical solutions to the Master equation for the wavefunction $\Psi[a,\phi]$ corresponding to expanding $(3+1)-dim$ worlds do not exist for the low initial energies $\Lambda$. Physically we can understand the out-of-equlibrum dynamics of vacua energies and massive modes on the landscape by thinking of all the matter modes contained in $\Sigma {\cal H}_n$ as making up a black hole with surface gravity $\kappa_{BH}$ and a DeSitter (inflationary) universe with vacuum energy $\Lambda$ and surface gravity of DeSitter geometry $\kappa_{DS}$. So, visually we have a black hole inside a DeSitter inflating space. The black hole is trying to pull the initial space into a point while the DeSitter geometry is trying to drive the universe into an accelerated expansion.  Depending which surface gravity is stronger determines whether the black hole 'pull' wins over the DeSitter drive to expansion or if the backreaction of the black hole on the geometry is just 'denting' an otherwise exponentially expanding space. The entropy $S$ of these out-of-equlibrium systems that are in an entangled state can be readily estimated from the Master equation. The general expression for $S$ is quite complicated but for the simplified case of the black hole inside a DeSitter space, the expression reduces to the entropy obtained for Schwarzschild-DeSitter geometries by \cite{gibbonshawking}, 

\begin{equation}
S \simeq (r_I -r_{f_{n}})^2 \,\,\,\,
r_I \simeq H_I^{-1} \,\,\,
r_{f_n} \simeq (H_I^{-3/2}<\phi_I\sqrt{U}>).
\label{desitterblackhole}
\end{equation}

where $r_I = l_H$ denotes the De-Sitter horizon of the inflationary patches with Hubble parameter $H_I$, entropy $S$, and $r_{f_N}$ the horizon of the 'black hole' made up from the $f_n$ modes where the average value of the mode $<f_n>$ is of order the mean vacuum state $\phi_I$,   $<f_n>\simeq \phi_I$.

The cuttoff derived for the ensemble of the initial states  above, namely, the lower bound on the survivor from backreaction of the universes, corresponds to $H_I =m, \phi_I = M_p$. It is interesting that this corresponds to the case of a zero entropy for the DeSitter - Black hole system, which is the case when the horizon and surface gravity $r_I^{-1}=\kappa_{DS}$ of the DeSitter patch coincide with the horizon and surface gravity $r_{ f_{n} }^{-1}=\kappa_{BH}$ of the black hole.
This means that a black hole with the same horizon as the initial inflationary patch is the borderline between the 'terminal' and 'survivor' universes. More to the point, the value of entropy $S=0$ provides the lower bound on the initial conditions $H_I,\phi_I$ for giving birth to a universe. 

To summarize: Initial states that start at high energies survive the backreaction of superhorizon modes on their space, thus they are named {\it 'survivor' universes}. Initial states that start at low energies can not survive the backreaction of the superhorizon sized 'environment' thereby developing gravitational instabilities that crunch them to a point. For this reason these are named the {\it 'terminal' universes}. Although we started with an ensemble of all possible initial conditions and energies on the string landscape multiverse, only a fraction of them, initial states with high energies $\Lambda$, are selected as physically relevant 'survivor' universes. 'Terminal' universes can not grow and therefore are not physically relevant since these initial states can not give birth to a universe. This new theory shows how the quantum dynamics of gravity and matter on a landscape multiverse results in a {\it superselection rule for the inital conditions of the universe}. The selection of 'survivor' universes that start at high energies and therefore low entropies is driven by the quantum dynamics of gravity in the ensemble of the initial conditions. Although the initial ensemble of all possible initial states has shrunk to a subset by being wiped clean of the low energy patches, it still contains a whole multiverse of high energy 'survivor' universes.

\section{Testable Predictions?}

Is this theory falsifable? This approach provides a satisfactory understanding of the deep mystery of the selection of the initial state, while offering a radically new way of thinking in terms of the quantum dynamics of gravity and the need for a multiverse framework. Thus the tantalizing question at this stage becomes: can the theory make predictions that can test it? As amazing as it sounds when considering that our framework is the multiverse, the possibility for hunting down imprints leftover from the earliest time and traces of the multiverse, is real.
The reasons are simple:

- the present universe provides a giant laboratory for testing microscopic phenomena that occured at its earliest times since the large scales today are a blown up version of the tiniest scales at the beginning, since they are redshifted from the small scales of the initial non-equlibrium phenomena;

- the sacred principle of quantum mechanics, our starting basis of the theory, is the Unitarity Principle,which states that information can never be lost. This means that if our universe started in a mixed state in the landscape multiverse, than it can never evolve into a pure state at late times. Information contained in the superhorizon size entanglement of our universe with all else outside can not be lost. This entanglement leaves its traces everywhere in the present observable sky.

These two reasons provide a whole series of imprints on the large scale structure (LSS) and cosmic microwave anisotropies (CMB) in the universe, derived in \cite{avatars1,avatars2}.

Finally, I should mention that if supersymmetry (SUSY) is a theory of nature than the landscape vacua energies were acquired by the SUSY breaking mechanism. These energies $\Lambda$ are then determined by the SUSY breaking scale '$b$' in them. This implies that there are two fundamental scales in the theory: the global scale which can be the Planck mass or string scale, $M_p$, and is  the same for all the vacua; and, the SUSY breaking scale $b$ which is a local scale since it varies from one vacuum state to the next. Whenever a theory contains two dimensionful constants of nature, these parameters and their combination have to be a part of the observables we measure. Thus, if our universe acquired its initial vacuum energy by breaking SUSY, this breaking scale will be found in our measurements.

How are these traces of entanglement from the very early universe calculated? We have a theory of Big Bang inflation that tells us that LSS and CMB were seeded by the primordial fluctuation. The perturbation theory also allows us to calculate the present structure and CMB in the universe from the primordial spectrum, through the evolution equations. The most important ingredient for the structure is the estimation of the background Newtonian gravitational potential $\Phi$ which maps the gravitational field of the large scale structure in the universe. We have direct measurements for $\Phi$. In the theory I have described here, besides the primordial fluctuation of inflation produced within our patch that seed the observed structure, we have a second channel of contribution to the Newtonian potential $\Phi$ for this structure which originates from the (highly non-local) superhorizon size entanglement of our universe with all the modes outside. The modification $\delta \Phi$ this nonlocal entanglement induces on $\Phi$,  can be derived in a precise manner by known physics equations and allows no room for phenomenology. When tracing out the effect of nonlocal entanglement on our universe, the energy of the initial waveapacket of our universe gets shifted and that shift produces the observable modification on the background potential of the universe. This modification to the primordial spectrum of perturbation and the Newtonian gravitational potential $\Phi$, derived in \cite{avatars1,avatars2},  which is superimposed to the standard inflationary potential $\Phi_0$, when evolved to present times carries all the observational imprints that can be observed in LSS and CMB, discussed in \cite{avatars1,avatars2}. 

The modification on $\Phi = [ \Phi_0 + \delta \Phi[b,V(\phi)]  ]$ originates from the backreaction term that was described in the previous section. Thus it has the opposite sign to the standard Newtonian potential $\Phi_0$, a case very similar to superimposing a negative mass on a positive mass. The result is an overall suppression of the {\it amplitude of perturbations, $\sigma_8$}, and a hole which in the sky show up as {\it voids}. But, considering that at present the strength of nonlocal entanglement is very small, then without resorting to the mathematical expression for $\Phi$ in \cite{avatars1,avatars2}, it can still be understood why the maximum modification, i.e. voids, will be found at very large scales, at the 'edges' of the universe, where the entanglement strength is relatively larger and physics is nearly linear. At shorter distances such as galaxy of cluster scales the messy nonlinear gastrophysics of structure formation would erase any small traces of modifications. We carried out this calculation for the modified Newtonian potential in \cite{avatars1,avatars2} and predicted that a giant void of a size of about $12$ degrees in the sky, should be found at about 8 billion lightyears away. Such a huge void can not be accommodated within the standard cosmology scenario which has only one channel for producing structure, a structure expected to be uniformly distributed by standard cosmology. Amazingly, this giant void was observed only a few months later, in 2007, by the Very Large Array Telescopes (VLA) radio measurements \cite{rudnick}. Although previously a cold spot in the skymaps of the first year and third year run of Wilkinson Microwave Anistropy Probe (WMAP) experiment had been spotted \cite{coldspot} the team were cautious in intrepreting the observation of nongaussianities as significant since the cold spot seemed to correspond to only a point-like source while the rest of the data in the skymaps were in excquisite agreement with inflationary predictions for a gaussian spectrum of the Microwave Background sky. A cold spot could, in principle, be taken as indirect evidence for the existence of a void. But it could also be consistent with the existence of textures, a possibility raised in $2007$ in \cite{turok}. Therefore the observation of the void made by Rudnick {\it et al} \cite{rudnick}, (see also [\ref{appendix}]), was highly significant because: unlike the indirect measurements of WMAP they found direct evidence for the void; and, because the void they observed happened to be at exactly the same location as the WMAP's cold spot, thereby confirming that the the nongaussianities seen in the WMAP's data were significant.

However this observed void is not the end in our void's 'story': As I mentioned, the largest effect where the modification to Newtonian potential $\Phi$ is at maximum strength, should be found at the largest present scales, the 'quadrupole' scales. These scales are separated by $60$ degrees in the sky which correspond to the earliest moments at the onset of inflation and the largest observable sky in the present universe. Our second testable predictio is that another void should be found at the 'quadrupole' scale.

The new channel of contribution originating from the nonlocal entanglement, and the modification it induces on our universe affects the expansion law and structure formations, can be understood by the fact that a modified Newtonian potential also results in a modified Friedmann equation for the expansion of the universe.

After including the backreaction due to the tracing of the long wavelength modes, the effective Friedmann equation was shown to be modified \cite{avatars1,avatars2} by a nonlocal term $F(b,V)$. This term reflects the nature of the entanglement of our universe, during Big Bang at high energies $\Lambda = V = m^2 \phi^2 /2$, and it also contains the second parameter of its SUSY breaking scale $b$:
\begin{equation}
\label{eq:modfried} H^2 = \frac{1}{3 M_{\rm P}^2}
\left[V(\phi)+\frac{1}{2} \left(\frac{V(\phi)}{3 M_{\rm
P}^2}\right)^2 F(b,V)\right]\equiv \frac{V_{\rm eff}}{3 M_{\rm P}^2}
\end{equation}
where
\begin{eqnarray} 
\label{eq:corrfactor}
F(b,V) &=& \frac{3}{2} \left(2+\frac{m^2M^2_{\rm P}}{V}\right)\log \left( \frac{b^2 M_{\rm P}^2}{V}\right)\nonumber \\
&-&\frac{1}{2} \left(1+\frac{m^2}{b^2}\right) \exp\left(-3\frac{b^2
M_{\rm P}^2}{V}\right). \end{eqnarray}
The scale at which the quantum nature of the nonlocal entanglement becomes powerful is given by something known as the interference length $L_1$, derived in \cite{avatars1}. This length can also be derived from the modification term $F(b,V)$.

 The traces of these modifications on LSS and CMB, due to the nonlocal entanglement, derived in \cite{avatars1,avatars2} agree with the data from WMAP \cite{wmap} and Sloan Digital Sky Survey (SDSS) \cite{sdss} experiments so far.For example, besides the void, the measurements for the amplitude of perturbation $\sigma_8$  by WMAP up to a few months ago, \cite{coldspot}, were in the range of values $\sigma_8 \simeq 1$, a value that was expected from inflationary scenarios. But the recently observed $\sigma_8$ by the 5-year WMAP run \cite{wmap} measures a suppressed value for the amplitude of fluctuations by a $30$ percent, $\sigma_8 \simeq 0.8$ in perfect agreement with our prediction for the suppressed amplitude of perturbations \cite{avatars2}.

But these modifications do depend sensitively on the second fundamental scale, that of SUSY breaking scale in our universe '$b$'. We dont know what the SUSY breaking scale is or at what energy scale inflation started. However it is clear that the modification term $F(b,V)$, which depends on $'b'$ was not strong enough to destroy the smoothness of the inflation potential energy (otherwise inflation would not have occured) for our survivor universe. Such a requirement introduces a lower bound on the strength of the modification and therefore on the SUSY breaking scale. On the other hand, through our CMB experiments that measure the effect of the entanglement channel on the inhomogeneities in the CMB,it  is constrained  to be of order $O[10^{-5}]$ up to distances a hundred times larger than our present horizon size $l_H = H_0^{-1}$. The latter observational constraint introduces an upper bound on the modification term and therefore the parameter $b$. A prediction of this theory, that will be tested by the Large Hadron Collider (LHC) at CERN soon, is that the SUSY breaking scale is bound by
\begin{equation}
\frac{V}{ M_p^{4}} < \frac{b^2}{ m^2} <  10^{-5}
\label{eq:general}
\end{equation}
 
We dont know at what energy scale $V$ inflation started. If our universe started with GUT scale inflation then the bounds on SUSY breaking scale for our universe are,

\begin{equation}
10^{-10} M_P < b < 10^{-8} M_P
\label{eq:susybounds}
\end{equation}

 which is a few orders of magnitude larger than the bounds expected by particle physics. 
These bounds will soon be tested by LHC at CERN and they can provide us with information on the energy scale $V$ of inflation.

So far I have advocated the need for extending our physical theories within the framework of a multiverse and made use of the string theory landscape as a working model for a physical multiverse. The multiverse field is becoming an important new direction of research in physics. But there are other types of multiverses besides the landscape multiverse. While the above approach can be applied to any type once its structure is known, physical reality can not correspond to more than one multiverse, or else we have simply shifted the mystery of the selection of our initial state to a bigger mystery, that of the selection of the multiverse. So, what is the Multiverse? In what spacetime does it exist? Little is known so far about this field. In the first part of my work, I dealt with a particular example, the string theory multiverse while 'brushing under the rug' key issues about the ontology of the multiverse, since the primary focus was on the puzzle of the selection of our initial state. I will now return to the topic of the basic questions about the structure and ontology of generic multiverses, that I ignored above.

\section{Multiverse Theories}

Three of our fundamental and most successful theories predict the existence of a multiverse. They are Quantum Mechanics, String Theory and the theory of Inflation, briefly described below. Currently the multiverse field is experiencing a renewal of interest and investigation. As I have argued recently, this may be due to the fact that the emergence of a multiverse from theories of nature is not a coincidence. Rather, predicting a multiverse may prove to become a requirement for the theory of quantum gravity. The motivation for defining the multiverse, along with the argument for needing to extend our conceptual framework to include the multiverse are discussed in Sec.5. In Sec.6, I conjecture the application of two principles in the multiverse that could help us define and investigate important questions for its ontology as well as providing a handle for testable predictions. Known theoretical examples of multiverses at present are: The Everett, Eternal Inflation, and, String Theory Landscape multiverses.

\subsection{ Everett Multiverse}
In the late 1950's Hugh Everett III gave his many worlds interpretation of Quantum Mechanics according to which, in the family of wavefunctions obtained from solving the Schrodinger equation, each wavefunction corresponds to a perfectly good physical reality (world). The Everett interpretation of Quantum Mechanics was complemented in the 1970's by H. Dieter Zeh \cite{zeh} and Wojciech Zurek \cite{zurek} who showed that the emergence of a classical world from the quantum wavefunction is achieved via a mechanism known as decoherence. Since then, the latter has been tested experimentally. The Everett multiverse, although debated for over 50 years, did provide the first scientific example of a multiverse being predicted from an underlying physical theory.

\subsection{ Eternal Inflation Multiverse}

The paradigm of inflation has been tremendously successful in explaining many of the observed features in our universe, such as flatness, homogeneity, structure and the scale-invariance of the cosmic microwave background (CMB) as well as being in nearly perfect agreement with the current observations so far \cite{wmap}. It should be noted however that inflation comes with a heavy price tag by creating a set of new puzzles, such as: why our universe started with such improbable initial conditions; why was the initial patch so smoothly fine-tuned; and, what is the inflaton? Some of these puzzles may appear less severe in the framework of eternal inflation. Current thinking has it that in many of the inflation models, eternal inflation is generic, i.e. once inflation starts it never entirely switches off. In every Hubble time during inflation, the volume of the inflating domain (Hubble volume) increases about 20 times. There is a nonzero probability that one of the domains, in the twenty newly created ones during one Hubble time, randomly ends up with a large enough fluctuation that can start inflation in that region, thereby producing a new bubble universe \cite{guth}. The conditions for such a large fluctuation to occur seem to be plausible for most inflationary models, thus etern inflation appears to be a generic prediction of the inflationary paradigm. This process of creating new bubbles via random large fluctuations, can continue {\it ad infinitum} into the future \cite{guth}, giving rise to an ensemble of bubble universes. If the universe continually reproduces new ones then the ensemble of all these bubbles provides a multiverse predicted by eternal inflation. The problem of the probability distribution of these bubbles in the eternal multiverse, known as the measure problem, remains an open issue. The weight assigned to the likelihood of having such large fluctuations and therefore eternal inflation, have been recently challenged and debated \cite{lauraparker,aguirre}.

\subsection{ String Landscape Multiverse}

Advances in String Theory over the last five years led to the discovery of the landscape of string theory \cite{landscape,ashok,douglas,bousso}. String theorists discovered that after reducing from higher dimensions to $(3+1)-D$ worlds, they did not end up with one unique solution but rather something like $10^{500}$ vacua solutions, the energy profile of which comprises the landscape. Work along this direction continues and it is possible that the number of string vacua solutions on the landscape, that contain $(3+1)-D$ worlds like our universe, will grow. In 2005, I proposed to view the ensemble of landscape vacua solutions as the string theory multiverse \cite{laurareview, land1} and argued that, since every vacua on the landscape is a potential birth place for a universe like ours, then the ensemble of these vacua is to be interpreted as a multiverse prediction of string theory \cite{laurareview, land1,land2,laurarich1}. An equivalent interpretation of the above proposal is to consider the string theory multiverse during its quantum phase, to be the physical phase space for the initial conditions \cite{land1,land2,laurarich1,laurarich2}. By proposing to place the wavefunction of the universe on the landscape and use its many-worlds interpretation \cite{land1,land2,laurareview}, the quantum mechanics multiverse was thus embedded onto the string landscape, thereby making the $Everett$ and $Landscape$ multiverses equivalent \cite{laurareview,laurarich1,laurarich2}.

\section{ Why do we need to live in a Multiverse?}

Breakthroughs in observational astrophysics have led us to an intriguing picture of the universe by challenging our previous understanding and expectations of nature.

Ten years ago we discovered that our universe is accelerating again, just like during the Big Bang inflation, but at lower energies. The acceleration is attributed to the most mysterious form of energy, dark energy. Since then, many other anomalies in the CMB have been spotted at the largest scales, scales compared to the horizon size. Such a lack of understanding of the present universe thus pushes the mystery of the selection of the initial conditions and inflation into the forefront of research. 

Naturally we expect that a deeper understanding of the mysteries of the birth of our universe and of its current acceleration, would stem from the theory of quantum gravity. But breakthroughs in the leading candidate for quantum gravity, string theory, led to a discovery of the landscape that is still considered by many to be a bizarre picture. The general view was that the multitude of vacua discovered presents a problem since we had expected that string theory would yield one unique solution corresponding to a universe that: started with high energy inflation; contained the correct values for the constants of nature; and, contained a tiny but nonzero amount of dark energy tuned exquisitely to $122$ orders of magnitude, which was just right for dominating the expansion of the universe at recent times. But string theory does not predict a unique universe, on the contrary, it predicts a multiverse$!$

A radically different point of view was advocated in \cite{laurareview,laurarich1,laurarich2}, namely: if we are (even allowed) to ask these deep questions about the selection of our universe then a multiverse picture must be an expected prediction for the theory of quantum gravity. How else can we meaningfully ask: why did we start with such an extremely improbable universe without implying as compared to what other possibilites?  Basically, a statement about an extraordinarily improbable event carries meaning only if we accept that other more probable events can be conceived. In this context then, the problem of the selection of the initial conditions implies that an extension of our theories to the multiverse framework is neither a philosophical contemplation, nor an unfortunate coincidence among our cherished theories, but could be a physical reality that is required in order to gain a deeper understanding of nature.

In fact, the extension of our current cosmology to a multiverse framework is an extension of the Copernician Principle to nature. Over the ages we found that we are not at the center of our solar system, that the solar system is not at the center of the universe and now, that the universe is not at the center of the world (the multiverse).

\section{What is the Multiverse: How Many Types are there?}

 If nature provided us with a multiverse, then a discussion of the ontology of the multiverse and the potential for extracting testable predictions, becomes of central importance in cosmology. Let us assume that the emergence of the background spacetime(s) has already been addressed by the underlying theory of quantum gravity, which also predicts the multiverse. Often in literature \cite{tegmark, davies, weinberg} the term 'multiverse' is often informally used to denote any multitude of universes. But by now we need a detailed discussion of issues such as their spacetime background, the relation between the different species, and ultimately their observational imprints. As described in Sec.1 we have already come across three seemingly unrelated examples of a multiverse. They can not all correspond to the same physical reality unless they are identical, because if they are distinguishable than we have simply transfered the puzzle of the selection of the initial conditions for our universe to a new puzzle, the selection of the 'real' multiverse. Thus the main question becomes: what is the multiverse, how many distinguishable species are there and, in what spacetime does the multiverse reside?

%%%%%%%%%%
Let us define the {\it 'multiverse'} to be the ensemble of all possible universes predicted by the underlying theory. For the sake of definiteness in terminology, consider the{\it ' universe'} to be (conservatively) defined as the domain of spacetime in which points were causally connected at some time slice, say $t=0$, of their past light cones. In an influential paper in 2003 \cite{tegmark} Max Tegmark classified the multiverses into four levels, (see also \cite{rudy}). Since then, as already discussed in Section 1, advances in string theory and precision cosmology have provided a wealth of information and, possibly observational imprints \cite{avatars1,avatars2,rudnick,wmap,coldspot} relevant to the multiverse. The previous discussion on the number of distinguishable multiverse species, their background space and the parameter of time can be viewed as an analysis of a multiverse corresponding to Tegmark's level $4$, (since his levels $1,2,3$ are subsets of level $4$), but with the addition of a new ingredient, namely the dynamics. Thus the correspondence of Level $4$ with the previous analysis is related to physical rather than intrepretational properties, in that the 'Multiverse' should be understood in terms of the structural hierarchy of Tegmark's Level $4$ that embeds all other types of multiverse levels within its domain. However the difference between the two definitions lies in the intrepretation of whether this embedding space is a physical or a mathematical entity. While I totally agree with Tegmark's view, (often referred to as a platonic view), that the physical existence of all mathematical solutions should be taken seriously, I would add dynamics as a new measure for weighting the mathematical solutions in this intrepretation. This means that only that subset of the initial mathematical entity which carries a dynamic description can be promoted to the multiverse corresponding to physical reality. I am thus proposing that we consider only multiverses that are equipped with a set of laws that allow their dynamical description, as candidates for the correspondence to physical reality, for the following reason: if this system has no dynamical evolution and description then its parameter of time becomes redundant. In this case a discussion of a background spacetime for the multiverse, or of our universe being part of it, do not carry any meaning. As we have seen, ignoring dynamics can prove quite dangerous in producing a set of paradoxes similar to those discussed in Section $1-4$ since the dynamics of the system, described by its own set of laws, is highly important in assigning and describing the physical properties of the system. A discussion of these physical properties and the background spacetime of a multiverse, which is the scope of the present study, can not be done within a static framework. Therefore, even if our starting point for a potential multiverse model is a mathematical structure, once we discover the dynamical laws of this entity and are able to provide a background spacetime framework, then the multiverse has been promoted to a physical entity.

It is therefore important to provide now a set of rules and an agreed taxonomy for the emerging field of multiverse physics. For simplicity let us discuss separately the different species of multiverses comprised from $(3+1)-D$ universes, and the types based on ensembles of universes with varying dimensionality. Within the multiverse, each domain can have:\\
A) its own set of laws;\\ 
B) its own set of constants of nature; or,\\
C) its own dimensionality\\
that vary from one domain to the next across the multiverse.

If our notion of a quantum to classical transition holds generically for some of the categories above, then just like the universe members, we can further fine-split the above multiverse species to two phases of existence, the quantum phase and the classical phase. The quantum multiverse is the phase space of the initial conditions for all members of the ensemble living in configuration space, whereas the classical multiverse is the ensemble of all the classical universes, embedded in a spacetime background, which are born from the quantum phase after the quantum to classical transition occurs, (perhaps similar to  the phenomenon of decoherence for our universe). It is possible that at some stage a mixed phase of quantum and classical universes may co-exist in the multiverse. I will focus only on the pure phases before the transition switches on and after the quantum to classical transition has been completed throughout the multiverse, and discuss below in a bit more detail the amount of information that can be extracted for each of the species ($A, B, C$), and in particular, the issue of the background space and time. 

%%%%%%%%%%

\subsection{ (Type A): Different laws across the Multiverse} 

In the type $A$ multiverse, by definition if we use the set of laws and equations that comprise our physical theories, we can not make predictions about the other domains since their physical laws are unknown to us and our theories are not relevant to them. These domains are beyond the realm of our understanding at present. 

What is time in the multiverse? Do we have the gravitational force in common with the other domains? Can gravitons from our domain leak across into the other sectors of the multiverse?  Many such issues remain open at present. Even if the answers to the latter question is yes which implies potential for observational imprints in our universe, we would still lack the computational power for making those predictions. It is possible that the multiverse predicted  by eternal inflation may fall into the type $A$ category with each bubble having its own spacetime, set of laws and maybe even different dimensionality. Since those laws are unknown to us and we lack an embedding theory for describing the varying laws of the bubbles ensemble, then the severe difficulties with the measure problem may be due to conceptual rather than technical issues.

\subsection{ (Type B): Same laws but different constants of nature across the Multiverse}

In the type $B$ multiverse, there definitely exists a force of gravity in all the domains since the laws and equations are the same as ours across this multiverse. In type $B$ gravity can spread everywhere and thus correlations among different domains do exist. In type $B$ multiverses we have the potential to make predictions about domains beyond our causal horizon. The Everett multiverse and the string theory landscape multiverse may be two such example of the type $B$ multiverses, since in both cases there is an overall embedding theory from which the ($3+1$)-D members of the ensemble (solutions) are derived, and which are closely related to each-other \cite{laurareview,laurarich1,land1,land2}.

\subsection{ (Type C ): Varying dimensionality Multiverse} This multiverse is the ensemble of universes comprised from domains have different dimensions. In general, any subset of Type $C$ that contain $(N+1)-D$ universes, ($N>3$), can be further be branched in the two groups $A^{N}$ and $B^{N}$ where $A^N, B^N$, correspond to the types $A, B$ defined above but in space dimensions $N>3$. This division is possible only if the parameter of time is fundamental in the type $ C$ multiverse. This argument makes it plausible for Type $A$ and Type $B$ multiverses to be sectors in the parent multiverse of Type $C$ since they could correspond to the sector of $C$ with $N=3$, where the higher $N$ dimensional branches are stacked above. 

Clearly a knowledge of what is time in the $C$ multiverse is crucial for its understanding and its relevance to our sector $N=3$.
Yet, since currently we lack a deeper understanding of this challenging issue, below I would like to propose that we apply the 'Principle of No Perpetum Mobile' for all multiverse species in order to achieve consistency in our guesstimates for their basic characteristics and embbedding spacetime.

\subsection{ Is Time in the Multiverse Fundamental or Emergent?\\   The Principle of 'No Perpetual Motion'}

Understanding the nature of time is another central mystery in theoretical physics. We have seen in Sec.2 that the question: is time fundamental or emergent, becomes crucially important to the multiverse physics. 

In Type $A$ multiverses it is possible to allow for time to emerge since the laws of physics vary across the multiverse. Specifically we can think of a Schrodinger equation which relates the parameter of time to the energy level of the wavefunction. If this equation is not valid for the whole ensemble then it is not difficult to envision a scenario where the parameter of time appears differently in various domains since it emerges from their different corresponding theories. We could also imagine that time acquires different meanings in different domains. In such a scenario it is plausible that the parameter of time may not appear as a fundamental quantity. This question becomes more severe for the type $C$ multiverse. 

In Type $B$ multiverses, since the laws of nature are the same everywhere, then the definition of the time parameter would also be the same. In this type, the existence of time as a fundamental property of nature appears to be the more plausible scenario.

Although there is no agreement yet on the issue of time being emergent or fundamental\cite{timearrow}, it is clear that
the nature of time and the multiverse are closely inter-related. Understanding one of these fundamental questions will shed light on the other and vice-versa.
Even though these problems remain open, based on the known symmetries and basic categorization of the multiverses, we can still derive some general statements about the parameter of time in the multiverse if we make the assumption that quantum mechanics is not an effective but a universal theory. If time translation is a symmetry of the multiverse (type $A,B$ or $C$) and the time-energy relation holds, then energy conservation resulting from this symmetry, implies that {\it 'no perpetual motion'} is allowed across the multiverse. At this stage of our knowledge we can not derive this conclusion since we don't know if the relation between time translation symmetry and energy conservation holds. 

Therefore I would like to propose that we impose the {\it The Principle of No Perpetual Motion in the Multiverse}, which requires that,(even if energy can be transferred), the energy can not be created or destroyed in or across domains in the multiverse. A by-product of this principle is to reduce the number of multiverse species that are candidates for physical reality to only those where the parameter of time does not vary across the multiverse, and is likely to be fundamental. This principle could reduce the number of possible multiverses corresponding to physical reality, to only those where the parameter of time is the same for all the domains, and possibly to narrow down the candidates to solely the multiverse where time appears fundamental.

\section{ In what spacetime does the Multiverse exists? \\  The Principle of 'Domains Correlations'}

Given that only one of the multiverses described above corresponds to the physical reality, and assuming that, independent of the space dimensionality, based on the principle of 'no perpetual motion', there is only one time parameter in that multiverse, then 
a crucial question is: {\it In what background spacetime does the multiverse exist?} 

This simple question is poorly understood at present. I would like to propose that we {\it use the 'Principle of Domains Correlations' in the multiverse as our criterion for determining the background spacetime(s)} in which the multiverse exists. According to this principle the existence of correlations among domains in the multiverse determines the background spacetime(s) on which the multiverse resides. Thereby the multiverse species fall in two broad distinguishable categories, $connected$ and $disconnected$ as follows:

{\bf Connected:} {\it If different domains in the multiverse are correlated then they must exist in the same background spacetime}.\\

{\bf Disconnected:} {\it If there are no correlations among domains in the multiverse, i.e. domains are totally disconnected from each other, then they must exist in different background spacetimes}.\\

The type $C$ multiverse could provide an illustration of the $Disconnected$ category unless the dimensionally different universes can all be embedded into a higher dimensional space into which correlations are carried via bulk space, as is the case in some brane-world scenarios \cite{randallsundrum} where gravitons spread in the bulk. If this is the case, then we have an example of a multiverse which 'lives' in higher dimensional space where these correlations exist although its member ('brane') universes belong to a lower space dimensionality.

The Connected Multiverse is the most interesting type, not only because it provides one embedding background space and observational handles but also because we have good indications, through CMB, that our universe may be a domain in one such type of the multiverse. Such was the case of the string theory multiverse discussed in \cite{avatars1,avatars2,rudnick}). As wasn shown in \cite{avatars1,avatars2} connectivity through the nonlocal entanglement of our domain with everything else on the multiverse left its imprints on the cosmic microwave background (CMB) and large scale structure (LSS), in various observables. Among them, it is worth mentioning: the prediction of a giant void \cite{avatars1} whose existence was later confirmed experimentally\cite{rudnick}; a suppressed $\sigma_8$ but an enhanced power with distinct signatures at higher multipoles in the power spectrum which is in agreement with the latest WMAP data release\cite{wmap}; and, a higher $SUSY$ breaking scale which will be tested this year by $LHC$. Thus the connected multiverse has the potential for providing observational clues. The Everett and Landscape multiverses, and in general the Type $B$ predictions fall under the 'Connected Multiverse' category.
If time is fundamental then a hierarchy of multiverses is possible with the Disconnected Multiverse type being the embedding space of all different sectors, such that Type $B$ is a sector of Type $A$, and type $B$ and $A$ are subsets of the larger multiverse, type $C$. 

The pressing question related to the variety of multiverse species investigated here is: {\it which one of them is real}? This question stands on shaky foundations since it can be answered only if we could define what constitutes physical reality. Although such definition does not yet exist, we all share the common notion that the existence of spacetime is part of the physical reality and that there can be but one reality. But having only one physical reality implies that only one of the multiverses can correspond to that physical reality. Following this argument seems to place the $Disconnected$ multiverse at a disadvantage. The lack of correlations among domains and the separate existence of sectors in different spacetimes means that there is no way or need for us to ever be aware of the other sectors existence in which case, for all relevance purposes, they are not part of {\it our} physical reality. If nature is economical then there is no need for the $Disconnected$ type which would make their existence unlikely. For the case of $Connected$ universes, due to the unitarity principle, those correlations will continue to exist at all times and all the sectors of this multiverse share the same background spacetime. Therefore all parts of the $Connected$ multiverse are relevant parts of reality for all times, which could make them a better candidate for describing nature.
   
Many of the issues discussed here raise more questions than provide answers. Yet we are on the first steps \cite{tegmark, laurareview, laurarich1,laurarich2} towards addressing a major problem, the ontology of the multiverse, that deserves deeper investigation.

\section{Thoughts on Defining the Multiverse}

Extending our physical theories to a multiverse framework may prove to be not only a neccesity for our understanding but also a fertile direction for exploring fundamental questions about or universe and nature. Progress in this field requires that by now we lay a set of principles and ground rules for discussing space and time in the multiverse and, ultimately reduce the class of candidates corresponding to physical reality to only one multiverse.

In this note I have proposed to apply two principles, that to my opinion, are required to discuss issues of space and time in which the classical multiverse is embedded and in which it is  assumed that quantum mechanics is a universal theory. The proposed principles for the multiverse are:

- the principle of {\it 'No Perpetual Motion',} which states that energy can not be created or destroyed in the multiverse; and,\\
- the principle of {\it 'Domains Correlations'} which states that if correlations among domains in the multiverse exists that there is only one corresponding spacetime for all universes.\\

Together these two principles help determine the background spacetime on which the multiverse is embedded.
Further, a possible existence of two phases, the quantum phase and the classical phase for the multiverse allows us to have a coherent picture for the member universes as they go through a quantum to classical transition and a dual picture in quantum Hilbert space and in real spacetime for the multiverse. These phases also allow us to discuss issues such as ergodicity of the phase space for the quantum multiverse and the possibility of Poincarre recurrences that in principle could continually create duplicates of each member universe.The phase space is defined as the ensemble of all the initial states, with each 'point' there representing the initial state of a possible universe. An ergodic phase space is one in which the volume of that space does not shrink. Just like an incompressable fluid, each 'point' on this phase space of initial conditions, our quantum multiverse, can flow along a trajectory. But as long as the total phase space remains incompressable, i.e. ergodic, this trajectory remains within the boundaries of the volume allowed to it. This means that if we wait long enough, the' point' will return to where it started, completing a Poincarre cycle and giving rise to a recurrence of the old state. But as was shown in \cite{laurarich1,laurarich2} an ergodic phase space for the quantum multiverse is highly unlikely due to the out-of-equilibrium dynamics of the degrees of freedom of each domain wipes out which much of the initial states in the phase space, thereby shrinking the volume to a subset of the original phase space, to the subspace corresponding to the 'survivor' universes 'points'. Since Poincarre recurrences are a consequence of ergodicity and the ergodicity of the phase space is broken then there is no danger that the universe could fluctuate back close enough to its previous state (thereby producing duplicates) no matter how long we wait for these Poincarre recurrences to occur\cite{laurarich2}. Dynamics does not allow temporary or eternal 'duplications'.

This paper attempts to touch upon some basic issues related to defining the multiverse and its correspondence to physical reality. Although more than $50$ years have elapsed since the first discussion of the 'many worlds' by Everett, we are in the process of resuming the first steps in setting the foundations and the ontology of the multiverse and of this new field in physics. In order to move forward at this relatively early stage in the field, we need to be clear on our set of principles and definitions, hence the need to open the discussion. Who would have though that Nature would lead us to a situation where a deeper understanding of its mysteries, at the smallest and largest scales, would guide the extension of our physical theories towards the realm of the multiverse?

\section{Appendix: Voids and the ISW effect}
\label{appendix}

Humans perform observations by receiving light signals. But light has to travel through the vast structures in the universe before reaching earth. These structures produce a gravitational field, with the field lines mapped by the Newtonian gravitational potential of structure, $\Phi$, as described in the text. Gravitational fields near superdense regions are strong and their lines are represented by deep 'wells' in $\Phi$, while the field is weak near the underdense regions and are represented by  'barriers' in the gravitational potential. Light emitted from an overdense region in the sky shows up as a hot spot in the primary CMB skymaps, and light from an underdense region shows up as a cold spot. But $\Phi$ changes with time due to the different energy components dominating the universe at various epochs, (except for the case when matter dominates the expansion). As light propagates through these changing structures in the universe, it can not escape the gravitational interaction with $\Phi$, an interaction which changes the energy of the photons.

Information on the time-dependent potential of structures is contained in the (late-times) Integrated Sachs-Wolfe (ISW) effect which provides a modification to the CMB spectrum, that is subdominant relative to the primary CMB spectrum. 
The ISW effect takes into account the total of the energy changes in the photons propagating through the time-dependent $\Phi$, from the (relatively late) time since the surface of the last scattering, (corresponding to about $1$ degree of the sky today,) to present time. When photons climb up and down deep potential wells then we receive a suppressed signal since it is harder for the photons to reach us, thus the signal is 'dampened' by such deep wells; when photons go through highly underdense or empty regions then there is no obstacle in their way thus the signal we receive is enhanced. But recently the universe has been dominated by dark energy. Dark energy makes the potential wells shallower by 'stretching' the fabric of spacetime. Therefore, due to the nonlinear ISW effect, we should observe an enhancement of the light signals reaching us, which is a consequence of the decay of the gravitational potential of large scale structure in the universe induced by dark energy. The signal is enhanced relative to the signal we would have received if the universe was matter dominated instead of dark energy dominated. For this reason and our knowledge of structure distribution, a detection of the ISW enhancement, (already observed by SDSS \cite{sdss}) is taken as evidence of dark energy.

But we should bear in mind that, although evidence for dark energy detection can be inferred from the ISW effect, this effect only measures enhancement or suppression of signals due to their propagation through the wells and barriers of $\Phi$. Therefore, the ISW effect provides 'direct' information only on the time-dependent profile of $\Phi$ and not on the factors that induce changes in this potential. A decaying $\Phi$ profile can be induced by the accelerated expansion which dilutes matter density, or be assigned to the existence of supervoids, or both. A void, (which shows up as a cold spot in the primary CMB \cite{coldspot}), by definition is a region of very low matter density and therefore, any signal received from the {\it void region will be enhanced - and not suppressed - due to the ISW effect}.
One example of a giant void was observed in $2007$ \cite{rudnick}, of about $12$ degrees in size and at a distance of about $8$ billion lightyears away, for which signals received should be enhanced from the ISW effect. But the CMB temperature observations \cite{wmap,coldspot} show a suppression of the signal at large distances, such as those of the void. This observation recently has given rise to a confusing intrepretation that the void \cite{rudnick} could be explained by the ISW effect by postulating that the suppresion of the CMB spectrum could be explained as a detection of the void's ISW effect. Au contraire, the ISW effect enhances the signal from voids. If we consider that the dominant component in our universe is dark energy and that the giant void exists, then the combined ISW signal enhancements from both dark energy and the giant void, are in fact incompatible with the observed suppressed signal of temperature anisotropies at large scales \cite{wmap,coldspot}. The combination of dark energy and the void, should show up as a larger than expected ISW enhancement, (an enhancement interestingly noticed in the data analysis of \cite{isw}). In short, a suppression of the signal at large scales due to the ISW effect does not provide an explanation for the void \cite{rudnick}, but rather makes the existence of giant voids even more unlikely and more difficult to explain.

%\maketitle

%\section{Introduction}

%%%%%%%%%%%%%NewIntroduction%%%%%%%%%%%%%%%%%%%%%%%%%%%

\section { Acknowledgements} 
%\begin{theacknowledgments}

LMH was suported in part by DOE grant DE-FG02-06ER41418, NSF grant PHY-0553312 and a FQXI grant.
%\end{theacknowledgments}

\end{document}